\documentclass{pasj00}
\draft

\begin{document}
\SetRunningHead{Machida & Matsumoto}
 {Dependence of Accretion Disk Temperature}
\Received{2000/12/31}
\Accepted{2001/01/01}

\title{Excitation of Low-Frequency QPOs in 
 Black Hole Accretion Flows }

\author{Mami \textsc{Machida}}%
\affil{Division of Theoretical Astronomy, 
National Astronomical Observatory of Japan, \\
2--21--1 Osawa, Mitaka-shi, Tokyo 181--8588}
\email{mami@th.nao.ac.jp}

\and
\author{Ryoji {\sc Matsumoto}}
\affil{Department of Physics, Faculty of Science, Chiba University, \\
1-33 Yayoi-cho, Inage-ku, Chiba 263-8522}
\email{matumoto@astro.s.chiba-u.ac.jp}

%


\newcommand{\rs}{r_{\rm s}}

\maketitle

\begin{abstract}

We present the results of global three dimensional magneto-hydrodynamic 
simulations of black hole accretion flows. 
We focus on the dependence of numerical results 
on the gas temperature $T_{\rm out}$ supplied from the outer region. 
General relativistic effects are taken into account using the 
pseudo-Newtonian potential. 
We ignore the 
radiative cooling of the accreting gas. 
The initial state is a torus whose density maximum is at 
$35 r_{\rm s}$ or $50 r_{\rm s}$ from the gravitating center, 
where $r_{\rm s}$ is the Schwarzschild radius. 
The torus is initially threaded by a weak azimuthal magnetic field.
We found that mass accretion rate and the mass outflow rate strongly 
depend on the temperature of the initial torus.  
The ratio of the average Maxwell stress generated by 
the magneto-rotational instability (MRI) to gas pressure, 
$\alpha \equiv \langle B_{\varpi} B_{\varphi}/4 \pi \rangle
/ \langle  P \rangle $ 
is $\alpha \sim 0.05$ in the hot torus 
($T_{\rm out} \sim 9.5 \times 10^9{\rm K}$ at $50 r_{\rm s}$) 
and $\alpha \sim 0.01$ in the cool torus 
($T_{\rm out} \sim 1.1 \times 10^9 {\rm K}$ at $35 r_{\rm s}$). 
In the cool model, 
a constant angular momentum inner torus is formed around 
$4-8 r_s$. 
This inner torus deforms itself from a circle to a crescent 
quasi-periodically.  
During this deformation, the mass accretion rate, the magnetic energy 
and the Maxwell stress increase. 
As the magnetic energy is released, 
the inner torus returns to a circular shape 
and starts the next cycle.

Power spectral density (PSD) of the time variation of the 
mass accretion rate in the cool model 
has a low frequency peak around $10 {\rm Hz}$ 
when we assumed a 10$M_{\odot}$ black hole. 
The PSD of the hot model is flat in $1-30 {\rm Hz}$. 
The slope of the PSD in the cool model is steeper than that in 
the hot model in $30-100 {\rm Hz}$.
The mass outflow rate in the low temperature model also shows 
quasi-periodic oscillation. 
Intermittent outflows are created in both models. 
The outflow speed is $0.01c - 0.05c$. 
The mass outflow rate is comparable to the mass accretion rate 
$\dot{M}$ at $2.5 r_{\rm s}$ 
in the high temperature model and about $0.02 \dot{M}$ 
in the low temperature model. 

\end{abstract}

\section{Introduction}

Accretion disks are believed to be the energy source of 
various activities in active galactic nuclei, 
X-ray binaries, protostars and so on. 
Using the RXTE satellite, detailed timing analysis of X-ray binaries 
has been carried out 
(e.g., \cite{hom2005a}, \cite{mcc2006}, \cite{sha2006}). 
Black hole candidates (BHCs) show transitions between a low/hard state (LHS) 
dominated by the hard power-law component and a high/soft state (HSS) 
characterized by the soft black body component. 
During the transition, BHCs stay in hard intermediate state (HIMS) 
or in a soft intermediate state (SIMS) \citep{hom2005b}. 
The light curves during LHS of BHCs are 
subject to violent X-ray fluctuations and 
sometimes accompany quasi-periodic oscillations (QPO) in the 
Fourier Power Spectral Density (PSD). 
The LHS is associated with steady outflows (e.g., \cite{gal2003}). 
In this state, mass accretes to the black hole as an optically thin, 
advection-dominated accretion flow 
(e.g., \cite{ich1977}; \cite{nar1994}, \yearcite{nar1995}). 
The energy spectrum of the HIMS is softer than that in the LHS. 
Low frequency ($1-10 {\rm Hz}$) QPOs are observed in the luminous 
hard state and in the intermediate state \citep{mcc2006}. 
High frequency QPOs ($\sim 100 {\rm Hz}$) are sometimes observed in 
these states. 
The frequency of low-frequency QPOs in the intermediate state 
moves to higher frequency and the low-frequency QPOs disappear 
when the state changes to HSS (see \cite{bel2006}).

These observations indicate that QPOs are associated with the cooling 
of the disk. 
By carrying out global three-dimensional (3D) magneto-hydrodynamic (MHD) 
simulations including radiative cooling, 
\citet{mac2006} showed that when the accretion rate exceeds the 
limit for the onset of the cooling instability, 
the radiatively inefficient, 
optically thin disk transitions into a magnetically 
supported, cool, intermediate state. 
\citet{oda2007} constructed a steady model of such disks and 
showed that their luminosity can exceed $0.1 L_{\rm Edd}$, 
where $L_{\rm Edd}$ is the Eddington luminosity. 
When the transition to the cool disk takes place in the 
outer region, 
cool gas will be supplied to the inner region.

Low frequency QPOs are sometimes attributed to the oscillation at 
the interface between the hot inner disk and the cool outer disk
(e.g., \cite{dim1999}). 
The excitation mechanism of such an oscillation, however, 
was not clear. 
Another puzzle is the coexistence of low-frequency and 
high-frequency QPOs. 
\citet{abr2001} proposed that high-frequency QPOs are generated 
by the resonance between radial and vertical oscillations of 
accretion disks.
S. Kato (2001a,b) pointed out that non-axisymmetric g-mode oscillations 
can be trapped in a thin, relativistic disk and 
these oscillations are excited by the corotation resonance. 
Non-linear couplings of disk oscillations and disk warp were 
examined by \citet{kats2004}.

\citet{katy2004b} reproduced high-frequency QPOs by 
3D MHD simulations of optically thin, geometrically thick 
accretion flows. 
They showed that the QPOs appear around the epicyclic frequency 
$\kappa$ at $4r_{\rm s}$ and 
$\Omega + \kappa$, where $\Omega$ is the Kepler frequency 
and showed that the amplitude of QPOs are damped after a while. 

The $1/f$-noise-like fluctuations observed in BHCs have been 
reproduced by global 3D MHD simulations (e.g., \cite{kaw2000}, 
\cite{haw2001}). 
\citet{kaw2000} showed that the PSD changes its slope 
around $10$ Hz when they assumed a $10 M_{\odot}$ black hole. 
\citet{mac2003} pointed out that  magnetic reconnection in the 
innermost region of the disk can be the origin of 
intermittent X-ray flares known as X-ray shots (\cite{neg1995}), 
which produce flat PSD at low-frequency ($\sim 1$ Hz).

In this paper, we report the results of global 3D MHD simulations 
which produced low-frequency QPOs and discuss their excitation 
mechanisms.

In section~\ref{method}, we describe basic equations and initial conditions. 
The results of simulations are given in section~\ref{result}. 
In section \ref{form}, we concentrate on the oscillation excited in the 
inner torus. 
Section~\ref{discuss} is devoted to discussion and conclusion.

\newpage

\section{Numerical Methods}  \label{method}

\subsection{Basic Equations}

We solved the following resistive MHD equations in a cylindrical 
coordinate system $(\varpi, \varphi, z)$; 

%
%
\begin{equation}
       \frac{\partial \rho}{\partial t} 
    +  \nabla \cdot ( \rho \mbox{\boldmath $v$} )
  = 
       0 ~,
\label{eqn:b1}
\end{equation}
%
%
\begin{equation}
      \rho \left[
      \frac{\partial \mbox{\boldmath $v$}}{\partial t}
    + \mbox{\boldmath $v$} \cdot \nabla \mbox{\boldmath $v$}
      \right]
  = 
    - \nabla P 
    - \rho \nabla \phi
    + \frac{\mbox{\boldmath $j$} \times \mbox{\boldmath $B$}}{c} ~,
\label{eqn:b2}
\end{equation}
%
%
\begin{equation}
     \frac{\partial \mbox{\boldmath $B$}}{\partial t}
  = 
     \nabla \times 
    ( \mbox{\boldmath $v$} \times \mbox{\boldmath $B$} 
    - \frac{4 \pi}{c} \eta \mbox{\boldmath $j$} ) ~,
\label{eqn:b3}
\end{equation}
%
%
\begin{equation}
     \rho T \frac{d S}{dt}
  = 
    \frac{4 \pi}{c^2} \eta j^2  ~,
\label{eqn:b4}
\end{equation}
where $\rho$, $P$, $\phi$, $\mbox{\boldmath $v$}$, $\mbox{\boldmath $B$}$, 
$\mbox{\boldmath $j$} = c \nabla \times \mbox{\boldmath $B$}/4 \pi$, 
$\eta$, $T$, and $S$ are the density, pressure, 
gravitational potential, velocity, magnetic field, current density, 
resistivity, temperature and specific entropy, respectively. 
The specific entropy is expressed as 
$S = C_{\rm v} {\rm ln} {(P/ \rho^{\gamma})}$,  
where $C_{\rm v}$ is the specific heat capacity and 
$\gamma$ is the specific heat ratio. 
We included the Joule heating term but neglected the radiative cooling term 
in the energy equation. 
We assume the anomalous resistivity 
$\eta = \eta_0 [{\rm max} (v_{\rm d}/v_{\rm c}-1,0)]^2$ 
\citep{yok1994},  
where $v_{\rm d} \equiv j/\rho$ is the electron--ion drift speed 
and $v_{\rm c}$ is the threshold above which 
anomalous resistivity sets in. 

General relativistic effects are simulated using the 
pseudo-Newtonian potential $\phi = -GM/(r-r_{\rm s})$ \citep{pac1980}, 
where $G$ is the gravitational constant, $M$ is the mass of the 
black hole, $r=(\varpi^2 + z^2)^{1/2}$, and 
$r_{\rm s}$ is the Schwarzschild radius.
We neglect the self-gravity of the disk.

\subsection{Numerical Methods and Boundary Conditions}

We solved the resistive MHD equations 
using a modified Lax--Wendroff scheme \citep{rub1967} with an
artificial viscosity \citep{ric1967}.  

The units of length and velocity are the Schwarzschild radius $r_{\rm s}$ 
and the light speed $c$, respectively. 
The unit time is $t_0=r_{\rm s} c^{-1} = 10^{-4}{M/10M_{\odot}}~ {\rm s}$. 
The unit temperature is given by 
$T_0 = m_{\rm p} c^2 k_{\rm B}^{-1}= 1.1 \times 10^{13} ~ {\rm K}$, 
where $m_{\rm p}$ is the proton mass and 
$k_{\rm B}$ is the Boltzmann constant.

The number of grids is $(N_{\varpi}, N_{\varphi}, N_z) = (250, 64, 384)$.  
The grid size is $\Delta \varpi = \Delta z = 0.1 $ for 
$0 < \varpi/r_{\rm s} < 10$, and $ |z|/r_{\rm s} < 10 $. 
For model LT, 
we set the grid interval as follows; 
$\Delta \varpi_{\rm n}= {\rm min}(1.05\Delta \varpi_{\rm n-1},
\Delta \varpi_{\rm max})$ , 
$\Delta z_{\rm n}= {\rm min}(1.05\Delta z_{\rm n-1}, 
\Delta z_{\rm max})$, 
where $\Delta \varpi_{\rm max}=10 \Delta \varpi$ and 
$\Delta z_{\rm max}=10 \Delta z$. 
The outer boundaries at $\varpi = 132r_{\rm s}$ and 
at $z = \pm 70 \rs$ are free boundaries where waves can be transmitted.  
For model HT, we set $\Delta \varpi_{\rm max} = 20 \Delta \varpi$ and 
$\Delta z_{\rm max}= 100 \Delta z$. 
Therefore, the outer boundaries are located at $\varpi=230 r_{\rm s}$ 
and $z = \pm 170 r_{\rm s}$, respectively.
The grid size in the azimuthal direction is $\Delta \varphi = 2 \pi / 63$. 

We included the full circle $(0 \leq \varphi \leq 2 \pi)$ 
in the simulation region, and 
applied periodic boundary conditions in the azimuthal direction. 
An absorbing boundary condition is imposed  at 
$r = r_{\rm in} = 2 \rs$ 
by introducing a damping factor,  
\begin{equation}
       D
  =  
       0.1 \left(
       1.0 - \tanh{
       \frac{r - r_{\rm in} + 5 \Delta \varpi}{2 \Delta \varpi}}
       \right) ~.
\label{eqn:b6}
\end{equation}
The physical quantities 
$q = (\rho, \mbox{\boldmath $v$}, \mbox{\boldmath $B$}, P)$ 
inside $r = r_{\rm in}$ are re-evaluated by 
\begin{equation}
       q^{\rm new} 
 = 
       q - D(q - q_0) ~,
\label{eqn:b7}
\end{equation}
which means that the deviation from initial values $q_0$ is 
artificially reduced with damping rate $D$. 
Waves propagating inside $r=r_{\rm in}$ are absorbed in the transition 
region ($r_{\rm in}-5 \Delta \varpi < r < r_{\rm in}$).

\subsection{Initial Model}

The initial state of our simulation is an equilibrium torus threaded by 
a weak toroidal magnetic field. 
At the initial state, the torus is assumed to have a specific 
angular momentum, $L\propto \varpi^{a}$.  

The magnetic field distribution is determined 
according to \citet{oka1989}. 
By using the polytropic relation $P=K \rho^{\gamma}$ 
at the initial state and by assuming 
\begin{equation} 
 \beta = \frac{8 \pi P}{B_{\varphi}^2} 
       = \beta_{\rm b} \left( \frac{\varpi}{\varpi_{\rm b}}
       \right)^{-2(\gamma-1)} ~ ,
\end{equation}
where $\beta_{\rm b}$ is the initial plasma $\beta$ at the 
initial pressure maximum of the torus 
$(\varpi,z) = (\varpi_{\rm b},0)$, and $B_{\varphi}$ is the 
azimuthal magnetic field. 
We integrated the equation of motion into a potential form, 
\begin{equation}
      \Psi(\varpi,z) 
 = 
      \phi + \frac{L^2}{2 \varpi^2} + \frac{1}{\gamma-1}v_{\rm s}^2 
     + \frac{\gamma}{2(\gamma -1)}v_{\rm A}^2 = \Psi_{\rm b}  
 =  {\rm constant} ~ , 
\label{eqn:poten}
\end{equation}
where $v_{\rm s} = (\gamma P/\rho)^{1/2}$ is the sound speed, 
$v_{\rm A} = B_{\varphi}/(4 \pi \rho)^{1/2}$ is the Alfv{\'e}n speed, 
and $\Psi_{\rm b} = \Psi(\varpi_{\rm b},0)$.  
At $\varpi = \varpi_{\rm b}$, the rotation speed of the torus 
$L/\varpi_{\rm b}$ equals the Keplerian velocity. 
By using equation (\ref{eqn:poten}), we obtain the density distribution as 
\begin{equation}
      \rho 
 = 
      \rho_{\rm b} \left\{
         \frac{\max{ [\Psi_{\rm b} - \phi -L^2/(2\varpi^2),0]}}
              {K[\gamma/(\gamma-1)]
                [1+\beta_{\rm b}^{-1}\varpi^{2(\gamma-1)}
       /\varpi_{\rm b}^{2(\gamma-1)}]} 
       \right\}^{1/(\gamma -1)}      ~ ,   
\end{equation}
where $\rho_{\rm b}$ is the density at 
$(\varpi,z) = (\varpi_{\rm b},0)$. 
Outside the torus, we assumed a hot, isothermal ($T=T_{\rm halo}$) 
spherical halo. 
The density distribution of the halo is given by 
$\rho_{\rm h} = \rho_{\rm halo} 
\exp [-(\phi-\phi_{\rm b})/(k_{\rm B} T_{\rm halo})]$, 
where $\phi_{\rm b}$ is the gravitational potential at 
$(\varpi,z)=(\varpi_{\rm b},0)$.

In this paper, we report the results of simulations for two models.
Model HT assumes a hot outer torus with the sound speed 
$c_{\rm b} =0.029c$ at $\varpi_{\rm b}=50\rs$. 
In model HT, the torus has a constant specific angular momentum
($a=0$). 
Model LT is a cool disk model in which $c_{\rm b} = 0.01c$ at 
$\varpi_{\rm b} = 35 \rs$ and $a=0.43$. 
In both models, we adopted  
$\beta_{\rm b} = 100$, $\gamma = 5/3$, $L=(\varpi_{\rm b}/2)^{1/2}
\varpi_{\rm b} /(\varpi_{\rm b}-1) \varpi^{a}$, 
$\rho_{\rm halo} = 10^{-4} \rho_{\rm b}$, 
$\eta_0 = 5 \times 10^{-4}$, and $v_{\rm c}= 0.9c$. 
Since we do not include radiative cooling, 
$\rho_{\rm b}$ is arbitrary. 
We adopt $\rho_{\rm b} = 1$.

\section{Numerical Results} \label{result}
\subsection{A hot accretion disk : model HT}


\begin{figure}
 \begin{center}
   \FigureFile(120mm,90mm){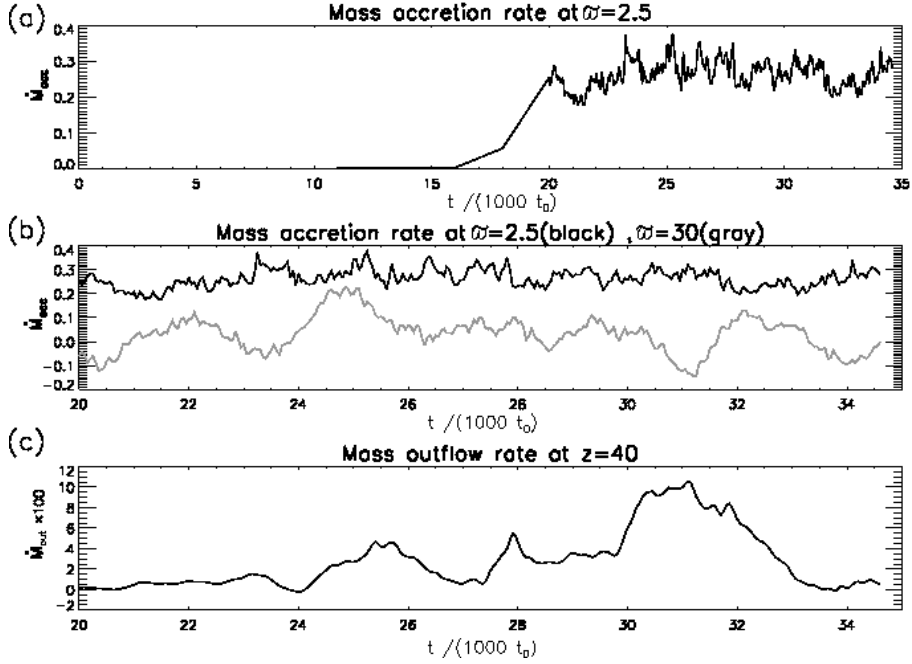}
 \end{center}
\caption{
(a) Time evolution of the mass accretion rate for model HT
measured at $\varpi=2.5 \rs$.
(b)Time evolution of the mass accretion rate at $\varpi=2.5 \rs$ 
(black) and $\varpi=30 \rs$ (gray). 
(c) Mass outflow rate measured at $z=40 \rs$. 
\label{fig:f1}}
\end{figure}

Figure \ref{fig:f1}a shows the time evolution of mass 
accretion rate for model HT measured at $\varpi=2.5 \rs$. 
Figure \ref{fig:f1}b shows the time evolution of mass 
accretion rate at $\varpi=2.5 \rs$ (black) and at $\varpi = 30 \rs$ (gray). 
The mass accretion rate $\dot{M}$ at $\varpi = 2.5 \rs$ is computed by 
\begin{equation}
 \dot{M} =\int_{-20}^{20}\int_0^{2 \pi} \rho \varpi v_{\varpi} 
d\varphi dz  ~ . 
\end{equation}
The mass accretion rate at $\varpi=30 \rs$ is measured by 
integrating the accretion rate in the equatorial region $|z|<5 \rs$. 
The unit of the mass accretion rate is 
$\dot{M}_{\rm 0}=2 \pi \rho_{\rm b} r_{\rm s}^2 c$.  
Mass accretion rate at $\varpi = 2.5 \rs$ becomes quasi-steady 
after about 10 rotational periods 
at the initial density maximum. 
The increase in mass accretion rate saturates when 
$\dot{M} \sim 0.3 \dot{M}_{\rm 0}$. 
The mass accretion 
takes place due to the efficient angular momentum transport 
by Maxwell stress in an MRI-driven turbulent field. 
The equatorial mass accretion rate at $\varpi =30 \rs$ shows time variation 
with a timescale of $\sim 4000 t_0$. 
Figure \ref{fig:f1}c shows the time evolution of 
mass outflow rate measured at $z=40 \rs$ computed by 
\begin{equation}
\dot{M}_{\rm out} = \int_{2}^{43}\int_0^{2 \pi} \rho v_{\rm z} \varpi 
d\varphi d\varpi ~ .
\end{equation}
The mass outflow rate correlates with the mass accretion rate with 
a time lag of about $4000 t_0$. 


\begin{figure}
 \begin{center}
   \FigureFile(120mm,90mm){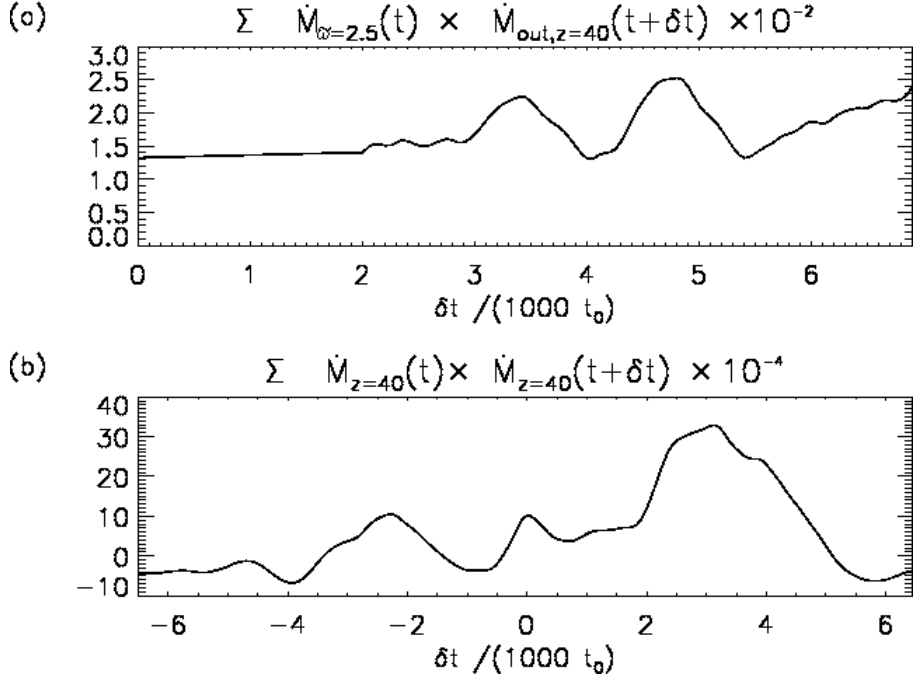}
 \end{center}
\caption{
(a) Correlation between the mass accretion rate at $\varpi=2.5\rs$ 
and the mass outflow rate at $z=40\rs$. 
(b) Self correlation of the mass outflow rate at $z=40 \rs$. 
\label{fig:f2}}
\end{figure}

Figure \ref{fig:f2}a shows the correlation between the mass 
accretion rate at $\varpi = 2.5 \rs$ and the mass outflow rate 
measured at $z = 40 \rs$ for model HT.
The correlation is computed by 
\begin{equation}
F(\delta t)=\int_{t_{\rm s}}^{t_{\rm e}} 
 [ (A(t) - \bar{A}) \cdot (B (t+\delta t)-\bar{B}) ] d t~.
\end{equation}
Here, $\bar{A}$ and $\bar{B}$ are average during the interval 
$t_{\rm s} < t < t_{\rm e}$. 
In figure 2, we adopted $t_{\rm s}=18000 t_0$ and $t_{\rm e}=25000 t_0$.
The correlation function has peaks at $\delta t = 3400 t_0$ and 
$\delta t = 4800 t_0$. 
The mass outflow rate at $z = 40\rs$ correlates with the mass 
accretion rate at $\varpi = 2.5 \rs$ with delay of 
$3000t_0 - 5000t_0$. 
This timescale is comparable to the time scale of the 
propagation time of the outflow whose average speed is $\sim 0.01c$. 
Figure \ref{fig:f2}b shows the self correlation function of the 
mass outflow rate at $z = 40 \rs$ in the time interval 
$21500 < t/t_0 < 33500$. 
In addition to the peak at $\delta t = 0$, positive peaks appear 
around $\delta t = -2400 t_0$ and $\delta t = 2400 t_0$. 
This indicates that mass outflow rate oscillates quasi-periodically 
with period $\sim 2400 t_0$.

\begin{figure}
 \begin{center}
   \FigureFile(120mm,90mm){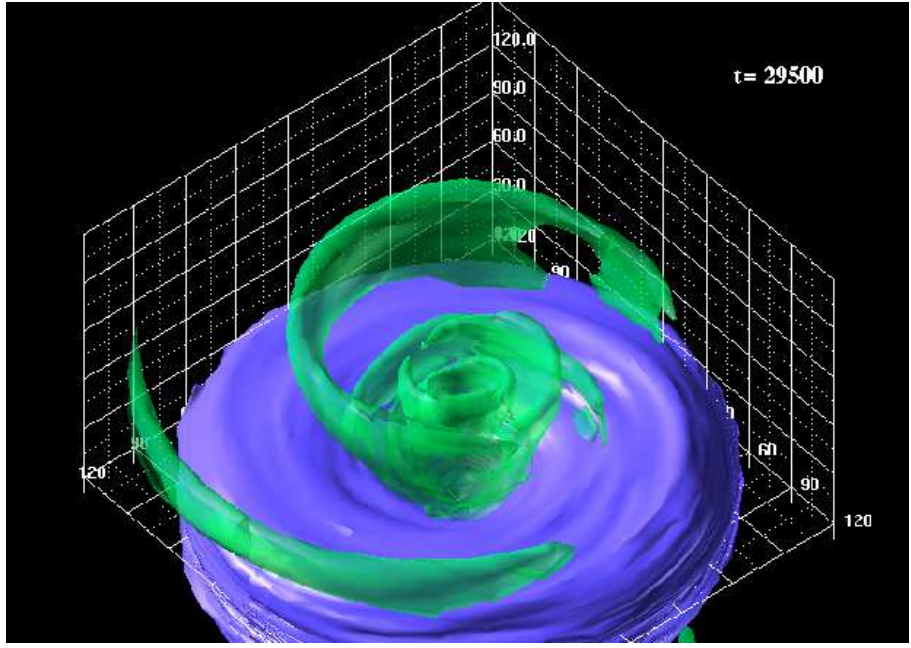}
 \end{center}
\caption{Emergence of outflows from the inner region of the disk. 
Blue surfaces show the isosurface of the density ($\rho = 0.2$) 
for model HT at $t=29500 t_0$. 
Green surfaces show the isosurface of the vertical velocity 
($v_{\rm z} = 0.05 c$). 
\label{fig:f3}}
\end{figure}

Figure \ref{fig:f3} shows the isosurface of the density and the 
vertical velocity. 
Blue surfaces and green surfaces depict the density 
isosurface ($\rho=0.2$) and the isosurface of vertical velocity 
($v_{\rm z}=0.05 c$), respectively. 
Winds emerge intermittently from the inner region of the accretion disk. 
Toroidal magnetic fields are dominant in the wind. 
The intermittent ejection is driven by expansion of magnetic loops 
anchored to the accretion disk \citep{katy2004}. 

\begin{figure}
 \begin{center}
   \FigureFile(120mm,120mm){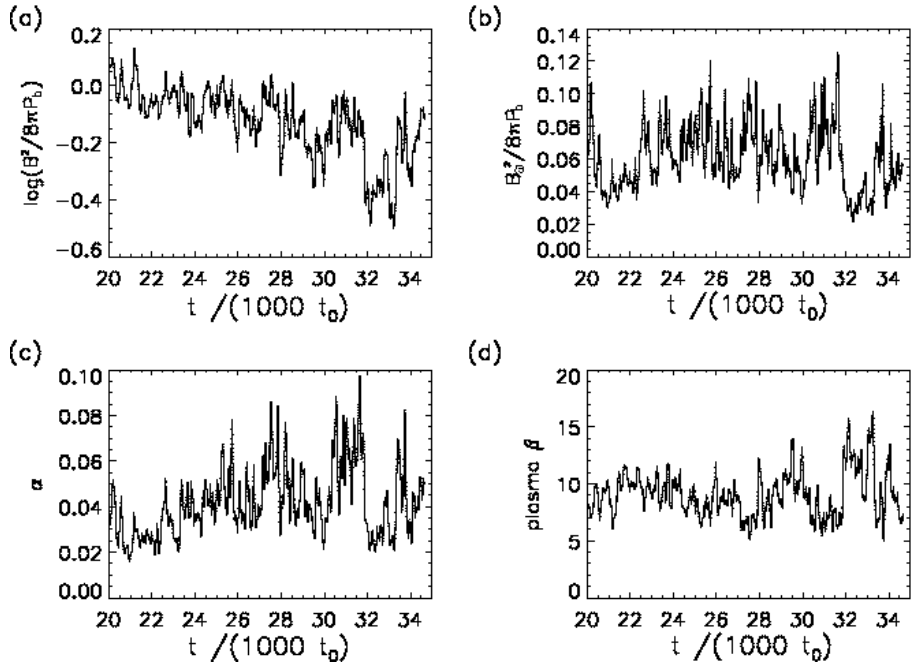}
 \end{center}
\caption{Time evolution of physical quantities for model HT 
averaged in $4 < \varpi/\rs <10$, $ |z|/\rs<1$, 
and $0 \le \varphi \le 2\pi$. 
(a) Magnetic energy, 
(b) $\langle B_{\varpi}^2/8\pi\rangle/ P_{\rm b}$, 
(c) time evolution of angular momentum transport rate, and 
(d) time evolution of the plasma $\beta$.
\label{fig:f4}}
\end{figure}

Figure \ref{fig:f4}a shows the time evolution of magnetic energy 
integrated in $4<\varpi/ \rs<10$, and $|z|/\rs<1$ normalized 
by the initial gas pressure at $(\varpi, z) = (\varpi_{\rm b}, 0)$. 
Figure \ref{fig:f4}b shows the time evolution of 
$B_{\varpi}^2/8\pi$ averaged in $4<\varpi/\rs<10$, and $|z|/\rs<1$. 
Figures \ref{fig:f4}c and \ref{fig:f4}d show the time evolution of 
$\alpha \equiv \langle B_{\varpi} B_{\varphi} / 4 \pi \rangle / 
\langle P \rangle$ 
and the ratio of the gas pressure to magnetic pressure 
$\beta \equiv P_{\rm gas}/P_{\rm mag}$, respectively. 
Although magnetic energy decreases, 
the plasma $\beta$ is nearly constant ($\beta \sim 8$)  
because gas pressure also decreases.   
The fluctuations of the magnetic energy and radial magnetic field correlate 
with the mass accretion rate. 

\subsection{A cool accretion disk: model LT}

\begin{figure}
 \begin{center}
   \FigureFile(120mm,90mm){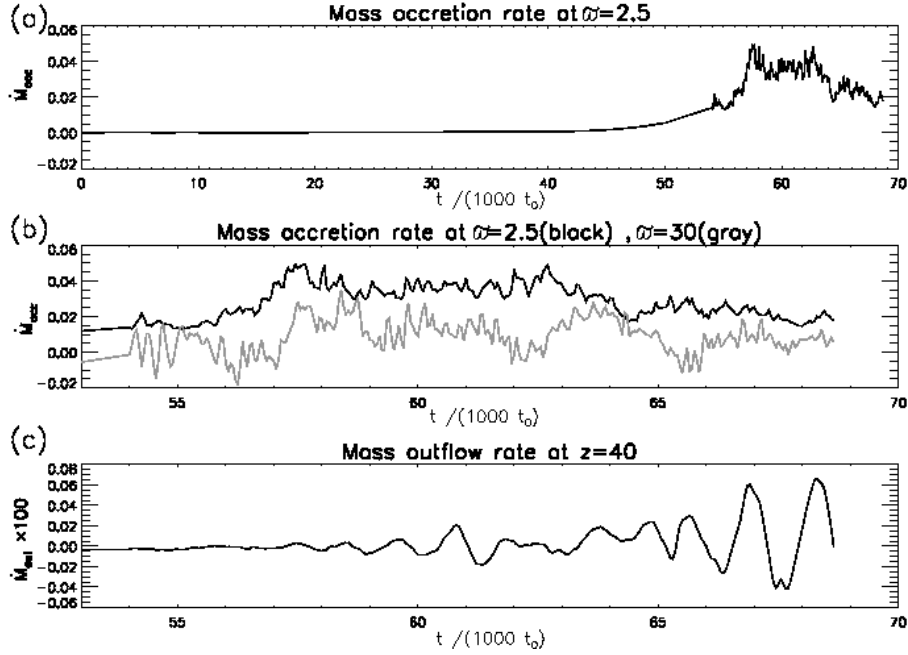}
 \end{center}
 \caption{
(a) Time development of the mass accretion rate $\dot{M}$ at 
$\varpi=2.5 \rs$ for model LT. 
(b)  Time evolution of the mass accretion rate $\dot{M}$ at 
$\varpi=2.5 \rs$ (black) and the equatorial mass accretion rate at 
$\varpi=30 \rs$ (gray). The interval $53000<t/t_0<70000$ is enlarged. 
(c) Mass outflow rate at $z=40 \rs$ for model LT. 
 \label{fig:f5}}
\end{figure}

Figure \ref{fig:f5}a shows the time evolution of the 
mass accretion rate for model LT measured at $\varpi=2.5\rs$. 
Figure \ref{fig:f5}b shows the mass accretion rate 
at $\varpi = 2.5\rs$ 
(black) and the equatorial mass accretion rate 
at $\varpi = 30\rs$ (gray) 
which enlarges the time range $53000 < t/t_0 < 70000$. 
Figure \ref{fig:f5}c shows the time evolution of the mass outflow rate 
measured at $z=40 \rs$. 
In Figure \ref{fig:f5}a, 
$\dot{M} \sim 0.04 \dot{M}_0$ is an order of magnitude 
smaller than that for model HT. 
This result indicates that the angular momentum transport rate 
strongly depends on the temperature of the gas supplied from 
the outer region.


\begin{figure}
 \begin{center}
   \FigureFile(120mm,90mm){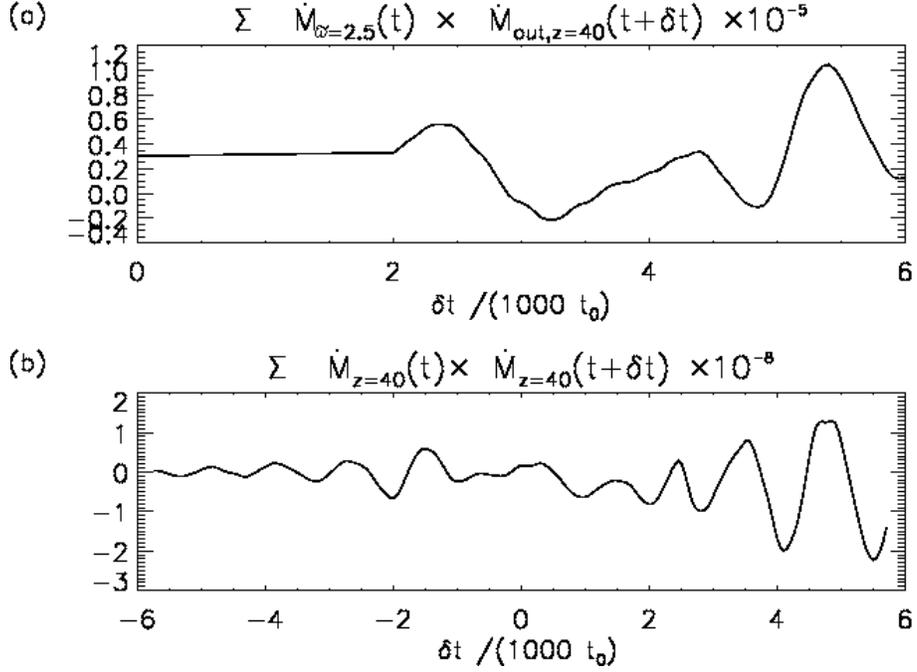}
 \end{center}
\caption{
(a) Correlation between the mass accretion rate at $\varpi=2.5\rs$ 
and mass outflow rate at $z=40\rs$. 
(b) Self correlation of the mass outflow rate at $z=40\rs$. 
\label{fig:f6}}
\end{figure}

Figure \ref{fig:f6}a shows the correlation between the 
mass accretion rate at $\varpi=2.5\rs$ and the mass outflow rate 
at $z=40\rs$ in the interval $52000 < t/t_0 < 62000$ 
for model LT. 
The mass outflow rate correlates with the mass accretion rate with 
time delay of $5000t_0 - 6000t_0$. 
Figure \ref{fig:f6}b shows the self correlation of the 
mass outflow rate at $z=40\rs$ in the interval 
$56650 < t/t_0 < 68650$. 
Positive peaks appear at $\delta t/t_0=-1500, 0, 2400, 3500$, and $4700$. 
It indicates that the mass outflow rate oscillates with 
period $1000t_0 - 2000 t_0$.

\begin{figure}
 \begin{center}
   \FigureFile(120mm,90mm){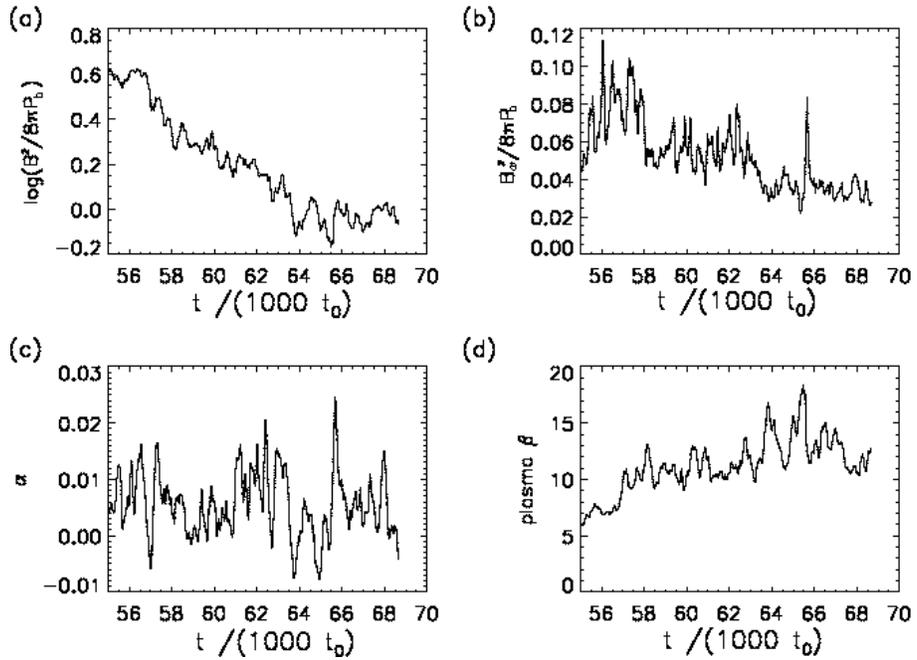}
 \end{center}
 \caption{
 Time evolution of (a) magnetic energy, (b) radial 
magnetic field, (c) angular momentum transport rate and 
(d) plasma $\beta$ for model LT. 
 \label{fig:f7}}
\end{figure}

Figure \ref{fig:f7} shows the time evolution of 
(a) magnetic energy, (b) $\langle B_{\varpi}^2 /8\pi\rangle/ P_{\rm b}$, 
(c) angular momentum transport rate $\alpha$, 
and (d) plasma $\beta$ for model LT.  
The magnetic energy is averaged in the region 
$4 < \varpi/\rs < 10$, $|z/\rs| < 1$, 
and $0 \le \varphi \le 2 \pi$.

Magnetic energy gradually decreases when $t > 55000t_0$. 
Figure \ref{fig:f7}c shows that $\alpha \sim 0.01$ in model LT. 
Since magnetic energy decreases, plasma $\beta$ increases and stays 
around $\beta \sim 10$. 
Physical quantities shown in Figure \ref{fig:f7} show short 
time scale oscillations and longer time scale ($\sim 4000 t_0$) 
time variations. 
The latter is due to the time variation of the mass accretion rate 
from the outer region.

\section{Formation of an Inner Torus and Its Oscillations} \label{form}

\begin{figure}
  \begin{center} 
    \FigureFile(120mm,120mm){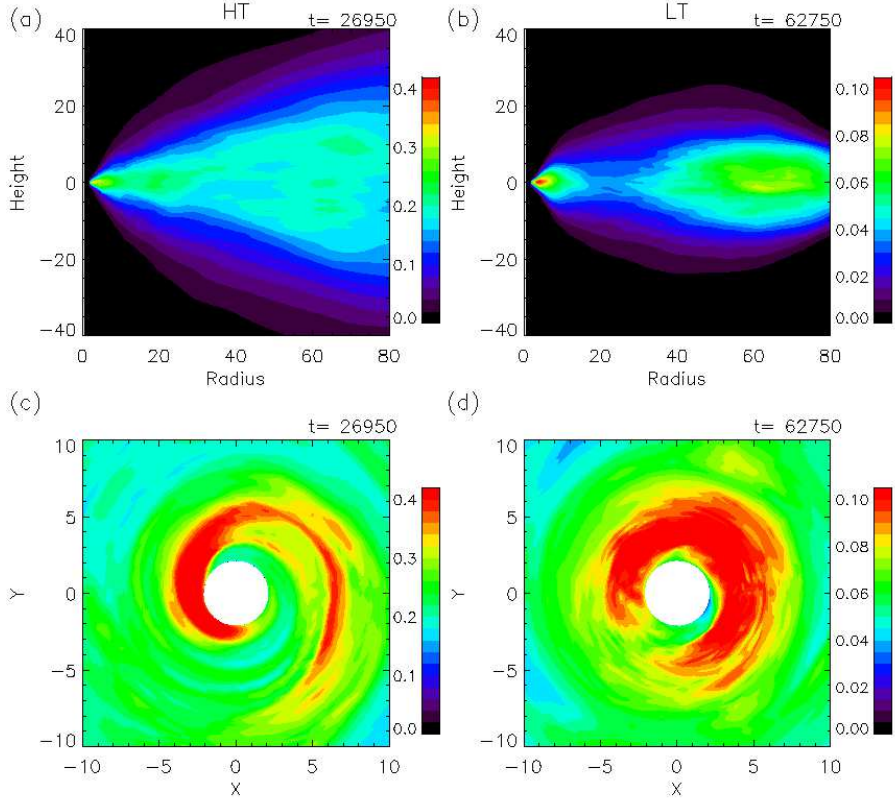}
  \end{center}
  \caption{
 The snapshots of the density distribution for model HT (left) 
and model LT (right). 
(a) and (b) show the density distribution 
averaged in the azimuthal direction. 
(c) and (d) show the density averaged in $|z|/\rs<1$. 
 \label{fig:f8}}
\end{figure}

Figure \ref{fig:f8} shows the snap shots of the density distribution. 
Figures \ref{fig:f8}a and \ref{fig:f8}b show the density distribution 
in $\varpi-z$ plane for model HT and LT, respectively. 
The density is averaged in the azimuthal direction. 
Figures \ref{fig:f8}c and \ref{fig:f8}d show the density distribution 
in $\varpi-\varphi$ plane averaged in $|z|/\rs<1$.  
In model LT, an inner torus is created around $\varpi/\rs \sim 4-8 $.
The inner torus is deformed into a crescent-like shape. 
The inner torus is formed because angular momentum transport becomes 
inefficient. 
Figures \ref{fig:f7}a and \ref{fig:f7}b indicate that 
Maxwell stress decreases due to the decrease in magnetic energy. 
Magnetic energy decreases partly because magnetic flux is swallowed 
into the black hole with accreting gas and partly because 
magnetic energy dissipates by magnetic reconnection. 
The deformation of the inner torus into a crescent shape takes place 
due to the growth of the Papaloizou-Pringle instability 
(\cite{pap1984}, \cite{dru1985}). 

When the disk is hot, since the inner torus is not formed, 
disk gas accretes to the black hole through dense, spiral 
channels (Figure \ref{fig:f8}c). 
This result is consistent with the result obtained by 
simulations of the hot disk reported by \citet{mac2003}.

\begin{figure}
  \begin{center}
     \FigureFile(120mm,90mm){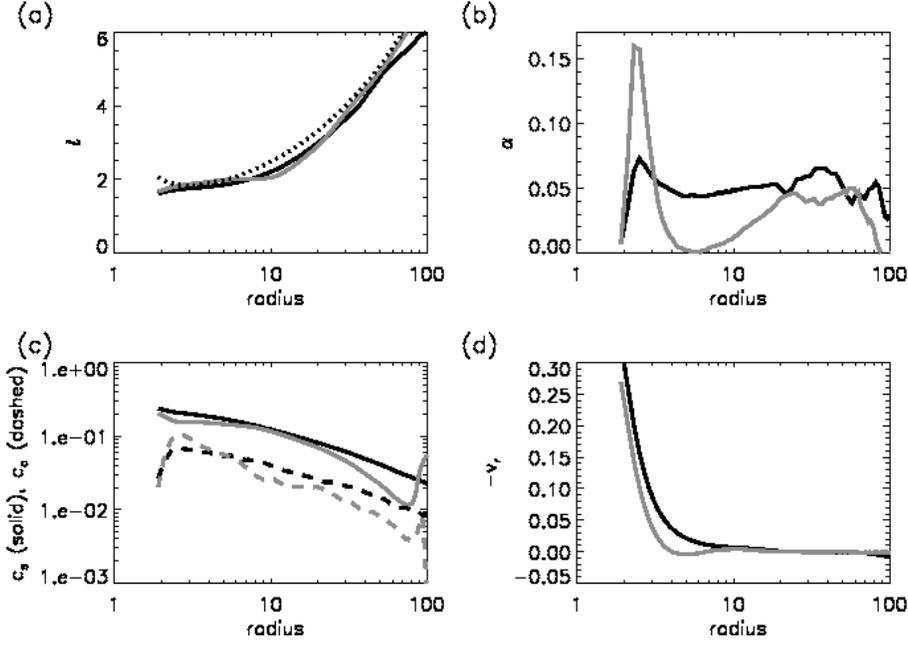}
  \end{center}
  \caption{ Radial distribution of physical quantities averaged in 
$|z|/\rs<1$ and in the azimuthal direction. 
Black curves show the distribution for model HT averaged in 
$23000 < t/t_0 <29000$. 
Gray curves show the distribution for model LT 
averaged in $55000 < t/t_0 < 61000$. 
(a) Time-averaged specific angular momentum distribution.
The dotted curve shows the Keplerian specific angular momentum 
using a Pseudo-Newtonian potential. 
(b) Radial distribution of the angular momentum transport rate 
defined by $ \alpha \equiv \langle B_{\varpi} B_{\varphi}/4 \pi \rangle 
/\langle P \rangle$. 
(c) Solid curves and dashed curves show the sound speed and 
Alfv\'en speed, respectively. 
(d) Distribution of radial velocity. 
  \label{fig:f9}}
\end{figure}

Figure \ref{fig:f9} shows the time-averaged radial distribution 
of physical quantities. 
The black curves show the results for model HT averaged in 
$23000<t/t_0<29000$. 
The gray curve shows the result for model LT averaged in 
$55000<t/t_0<61000$. 
Figure \ref{fig:f9}a displays the specific angular momentum distribution. 
The dotted curve shows the Keplerian specific angular momentum 
using pseudo-Newtonian potential.  
Since angular momentum continuously decreases in model HT, 
no inner torus is formed in this model. 
In model LT (dashed curve), since $\alpha$ is small, 
nearly constant angular momentum, 
inner torus is formed in $\varpi /\rs \sim 4-8$. 

Figure \ref{fig:f9}b shows the radial distribution of 
$\alpha \equiv \langle B_{\varpi} B_{\varphi}/4 \pi \rangle / 
 \langle P \rangle$. 
Inside the inner torus, 
the angular momentum transport rate becomes very small. 
Solid and dashed curves in Figure \ref{fig:f9}c show 
the sound speed and Alfv\'en speed, respectively. 
Figure \ref{fig:f9}d shows the radial velocity. 
The accretion proceeds subsonically in $\varpi > 3 r_{\rm s}$. 
In model LT, 
accretion speed becomes very low in the inner torus.

\begin{figure}
  \begin{center}
     \FigureFile(120mm,120mm){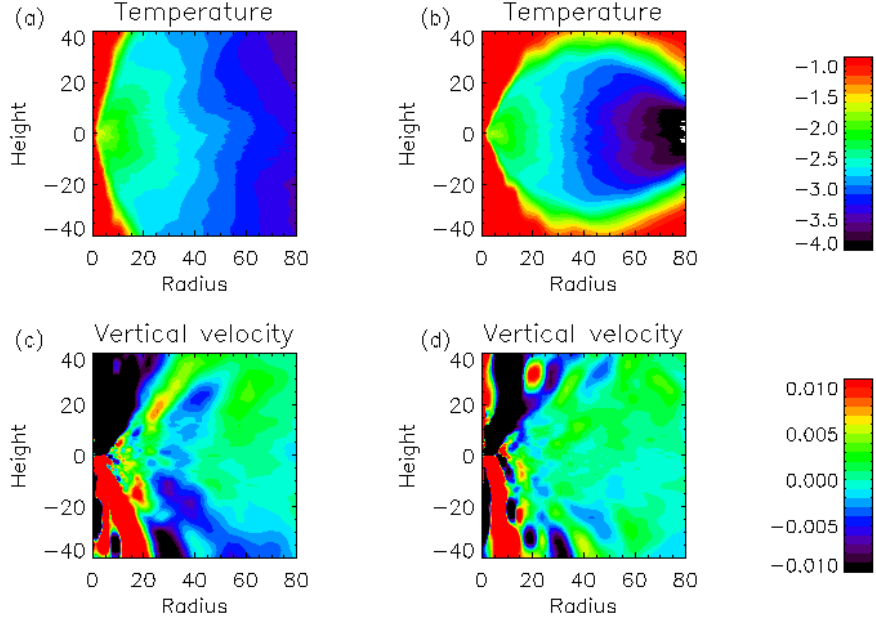}
  \end{center}
  \caption{(a) and (b) $\varpi - z$ slice of distribution 
of temperature $\log (T/T_0)$ averaged in the azimuthal direction. 
(c) and (d) distribution of  the vertical velocity 
averaged in the azimuthal direction.  
(a) and (c) are at $t = 26950 t_0$ for model HT, and 
(b) and (d) are at $t = 62750 t_0$ for model LT.
  \label{fig:f10}}
\end{figure}

Figure \ref{fig:f10} shows the distribution of azimuthally averaged 
temperature (\ref{fig:f10}a, \ref{fig:f10}b), 
and vertical velocity (\ref{fig:f10}c, \ref{fig:f10}d). 
Figures \ref{fig:f10}a and \ref{fig:f10}b show that 
accreting matter is significantly heated up. 
Figures \ref{fig:f10}c and \ref{fig:f10}d show that mass 
outflow emerges from the disk with an average speed of $0.01c$. 
The outflow is anti-symmetric to the equatorial plane. 
The outflow is more powerful in model HT. 
By comparing Figures \ref{fig:f8}a and \ref{fig:f10}a, 
we can distinguish three regions; 
equatorial disk, hot funnel near the rotation axis, 
and outflows between the disk and the funnel. 
Accretion proceeds in the equatorial disk. 
A fraction of the accreting matter is ejected from the 
inner region of the disk.

\begin{figure}
  \begin{center}
     \FigureFile(120mm,120mm){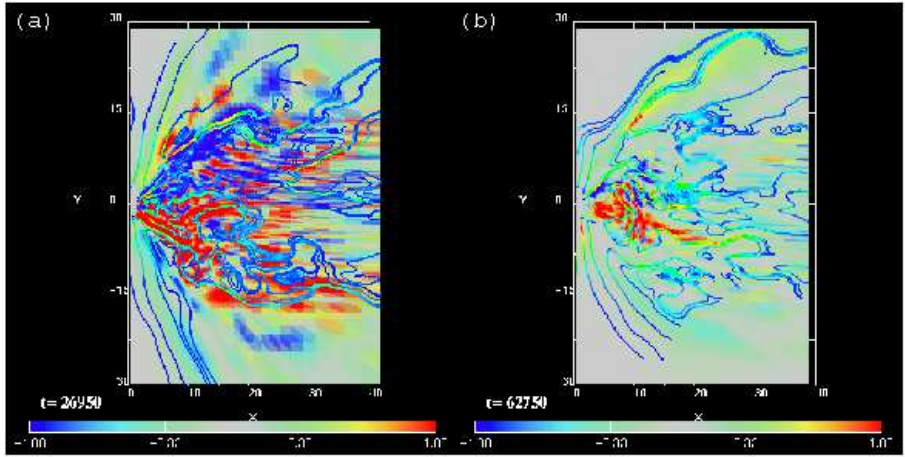}
  \end{center}
  \caption{ Distribution of azimuthal magnetic field (color) and 
magnetic field lines (blue curves). 
The magnetic field lines are depicted from the poloidal components of 
magnetic fields averaged in the azimuthal direction. 
(a) model HT and (b) model LT .
  \label{fig:f11}}
\end{figure}

Figure \ref{fig:f11} shows the distribution of the azimuthal magnetic 
field (color) and magnetic fields depicted from the poloidal 
components of magnetic fields averaged in the azimuthal 
direction (curves). 
The azimuthal magnetic field is antisymmetric to the equatorial plane 
at this stage ($t=26950 t_0$ for model HT 
and $t=62750 t_0$ for model LT). 
Magnetic fields are turbulent inside the disk 
but show more coherent structures in the interface between the 
disk and the halo where mass outflow takes place. 
Magnetic field lines are stretched along this interface. 
A large-scale poloidal magnetic field is created in the 
funnel near the rotation axis. 

\begin{figure}
  \begin{center}
     \FigureFile(120mm,120mm){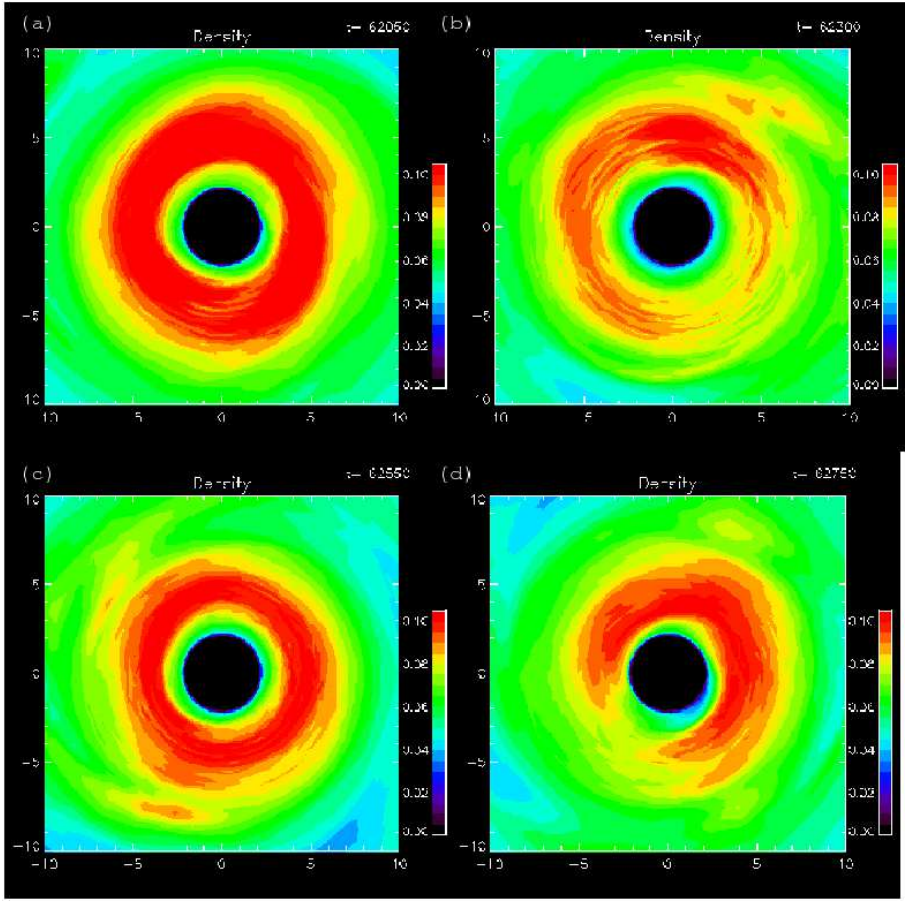}
  \end{center}
  \caption{ Snapshots of the density distribution on 
$\varpi-\varphi$ plane for model LT. 
  \label{fig:f12}}
\end{figure}

Figure \ref{fig:f12} shows the snapshots of the density 
distribution on the $\varpi - \varphi$ plane for model LT. 
Density is averaged in the same region as 
that in Figures \ref{fig:f8}c and 
\ref{fig:f8}d. 
The inner torus deforms its shape from a circle into a crescent, 
and from a crescent into a circle, repeatedly. 

\begin{figure}
  \begin{center}
     \FigureFile(120mm,120mm){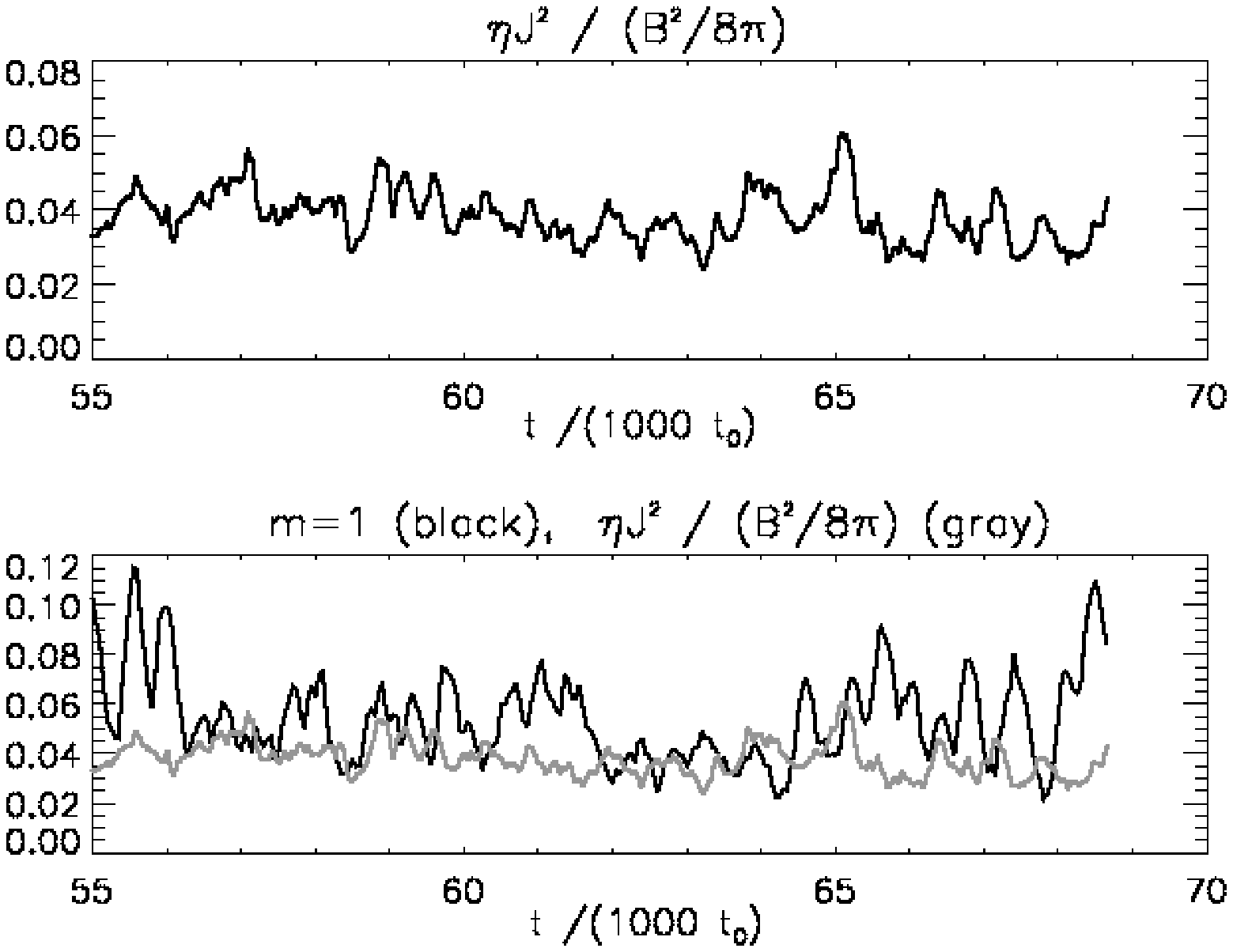}
  \end{center}
  \caption{ (top) Time evolution of the ratio of the Joule heating 
rate to the magnetic energy for model LT. 
(bottom) Time evolution of the Fourier amplitude of the 
non-axisymmetric mode with the azimuthal mode number $m=1$ computed 
from the density distribution (black). 
Gray curve depicts the same curve as that in the top panel. 
  \label{fig:f13}}
\end{figure}

The top panel of Figure \ref{fig:f13} plots the time 
evolution of the ratio of the Joule heating rate to the magnetic energy 
averaged in $4 < \varpi/\rs <10$, $|z|/\rs<1$, and $0 \le \varphi \le 2\pi$. 
The ratio increases when magnetic energy is released 
(when $\langle B^2/8\pi\rangle$ decreases or $\eta J^2$ increases). 
The ratio changes quasi-periodically with a time scale of $\sim 1000 t_0$.

The bottom panel of Figure \ref{fig:f13} shows the time evolution of 
the amplitude of non-axisymmetric $m=1$ mode 
($m$ is the mode number in the azimuthal direction) of the 
density. 
The amplitude of the azimuthal mode is computed by Fourier 
decomposing the density contrast 
\begin{equation}
\rho/\langle \rho \rangle = 
\rho/ \int_{2.9}^{5.3} \int_{-1}^1 \rho d \varpi dz  ~.
\end{equation}
The running average of the amplitude is plotted using the 
amplitude during $250 t_0$ for each point.
The gray curve depicts the same curve as that in Figure \ref{fig:f13}. 
The amplitude of $m=1$ mode anti-correlates with 
$\eta J^2/\langle B^2/8 \pi \rangle$. 
This indicates that the magnetic energy is released when the $m=1$ 
mode disappears. 
The amplitude of the $m=1$ mode also correlates with 
$\langle B_{\varpi}^2 \rangle$ (Figure \ref{fig:f7}b) and 
$\alpha$ (Figure \ref{fig:f7}c). 
They all show long timescale ($t \sim 4000 t_0$) variation 
and short timescale ($t \sim 1000 t_0$) oscillations.  
The peaks of the short timescale oscillations in the Fourier amplitude 
of the $m=1$ mode in Figure \ref{fig:f13} 
coincide with those in Figure \ref{fig:f7}.

\begin{figure}
  \begin{center}
     \FigureFile(120mm,120mm){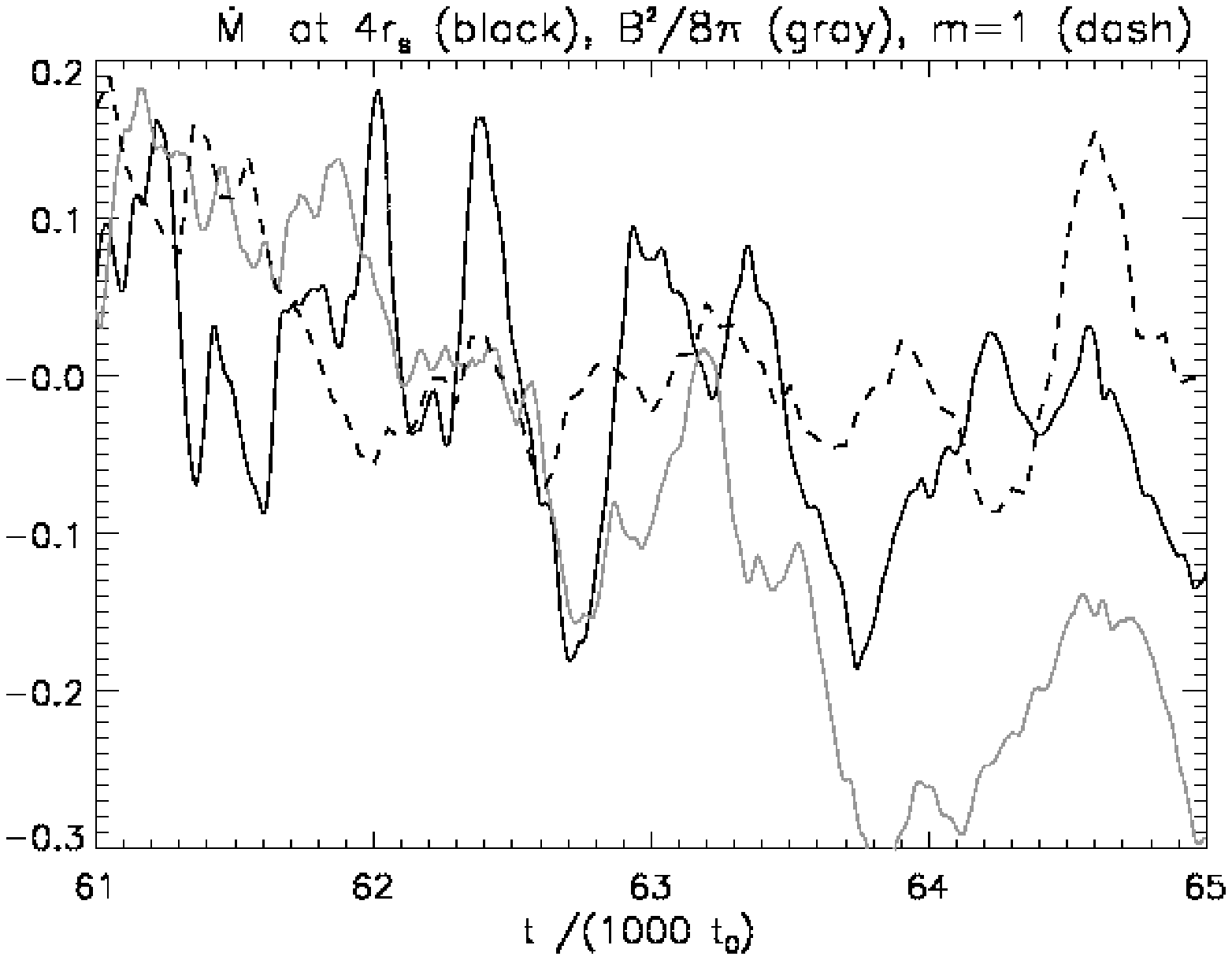}
  \end{center}
  \caption{ 
Correlation of the time evolution of the mass accretion rate and 
magnetic energy in model LT. 
(black) Time evolution of the mass accretion rate measured 
at $\varpi = 4 \rs$. 
(gray) Time evolution of the magnetic energy integrated in 
$4 < \varpi/\rs < 10$, $|z|/\rs<1$, and $0 \le \varphi \le 2 \pi$. 
(dashed) Fourier amplitude of the $m=1$ mode in the density distribution. 
The curves are arbitrarily shifted in the vertical direction. 
  \label{fig:f14}}
\end{figure}

Figure \ref{fig:f14} shows the 
mass accretion rate measured at 
$\varpi = 4\rs$ (black), magnetic energy (gray),  
and the Fourier amplitude of the $m=1$ mode for the density 
(dashed) for model LT. 
When the magnetic energy increases inside 
the inner torus,  the mass accretion rate increases 
because the angular momentum transport rate increases as 
magnetic energy (and magnetic stress) is accumulated in the torus. 
On the other hand, when the magnetic energy is released, 
the mass accretion rate decreases. 
The amplitude of the $m=1$ mode (dashed curve in Figure \ref{fig:f14}) 
correlates with the magnetic energy (gray). 
The magnetic fields are amplified when the amplitude of the 
$m=1$ mode increases. 
On the other hand, the amplitude of the $m=1$ mode decreases when 
the magnetic energy is released. 
When the Maxwell stress decreases due to the decrease in 
magnetic energy, the Papaloizou \& Pringle instability grows 
again inside the torus. 
Therefore, the torus deforms itself into a crescent shape. 
The interval between the magnetic energy releases in the inner torus 
is $\sim 1000 t_0$.

\begin{figure}
 \begin{center}
   \FigureFile(120mm,90mm){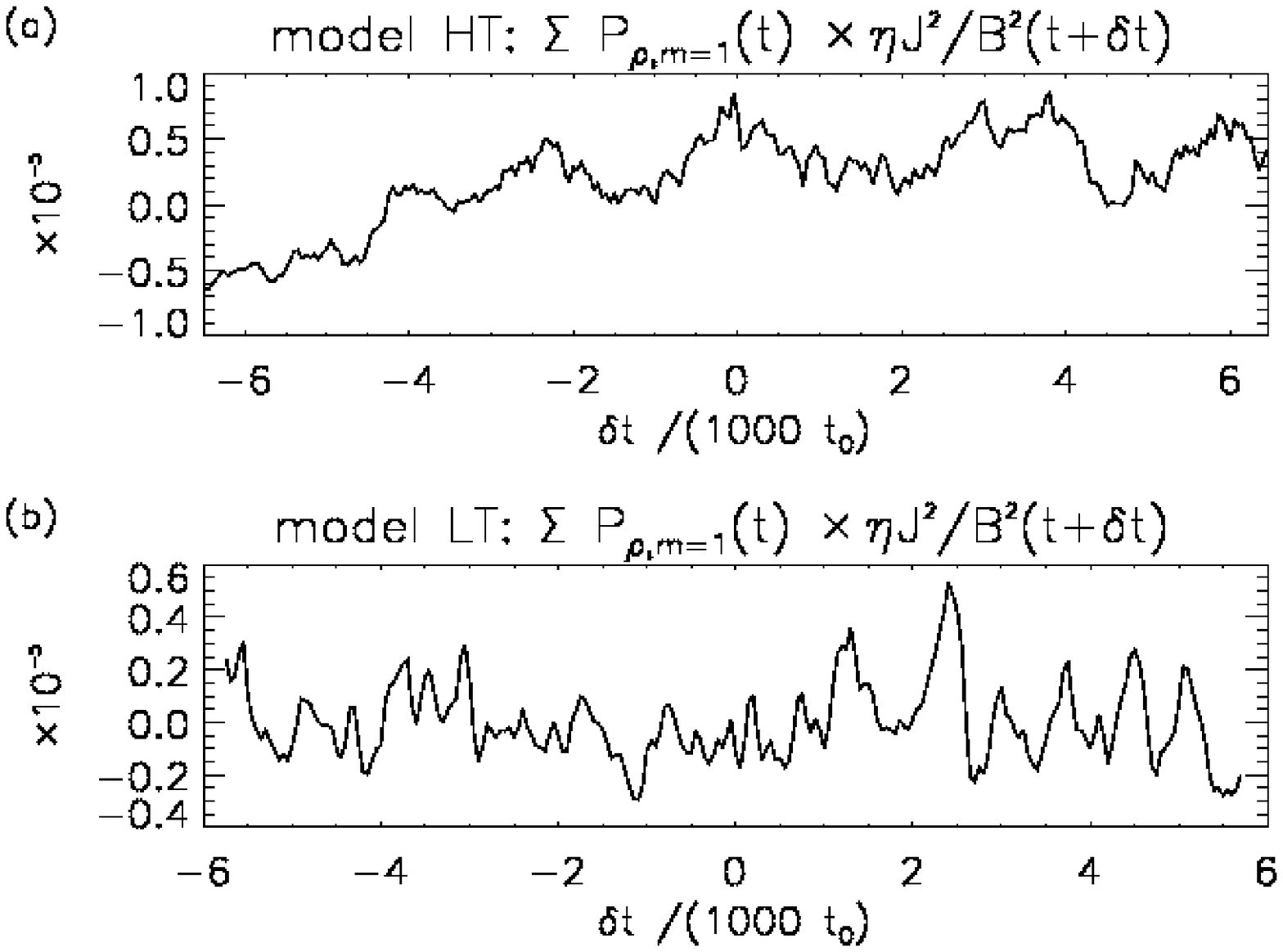}
 \end{center}
 \caption{
Correlation function between the amplitude of the $m=1$ mode 
and the Joule heating rate normalized by the magnetic energy. 
 (a) is model HT and (b) shows model LT.
 \label{fig:f15}}
\end{figure}

Figure \ref{fig:f15}a shows the correlation between the 
amplitude of the $m=1$ mode and the Joule heating rate 
normalized by the magnetic energy for model HT. 
The correlation function has positive peaks at 
$\delta t/t_0=-2500, 0$ and $3000$. 
The positive correlation at $\delta t=0$ indicates that 
magnetic energy is released when $m=1$ spiral channel 
(see Figure \ref{fig:f8}c) develops. 
This result is consistent with that reported in 
\citet{mac2003}, where we showed that magnetic reconnection 
takes place in the spiral channel. 
Other peaks in the correlation function indicate that 
$m=1$ mode develops quasi-periodically with 
interval $2500 - 3000 t_0$. 

Figure \ref{fig:f15}b shows the correlation between the 
amplitude of the $m=1$ mode and the Joule heating rate for 
model LT. 
In contrast to that in model HT, 
no positive peak appears at $\delta t=0$. 
Instead, negative peak appears at $\delta t/t_0=-1200$ and 
positive peaks appear at $\delta t/t_0 = 1200, 2400$. 
It indicates that $m=1$ mode anti-correlates with 
the Joule heating rate and 
that the growth of the $m=1$ mode precedes the release of the 
magnetic energy by $\delta t \sim 1200 t_0$. 

\begin{figure}
 \begin{center}
   \FigureFile(120mm,90mm){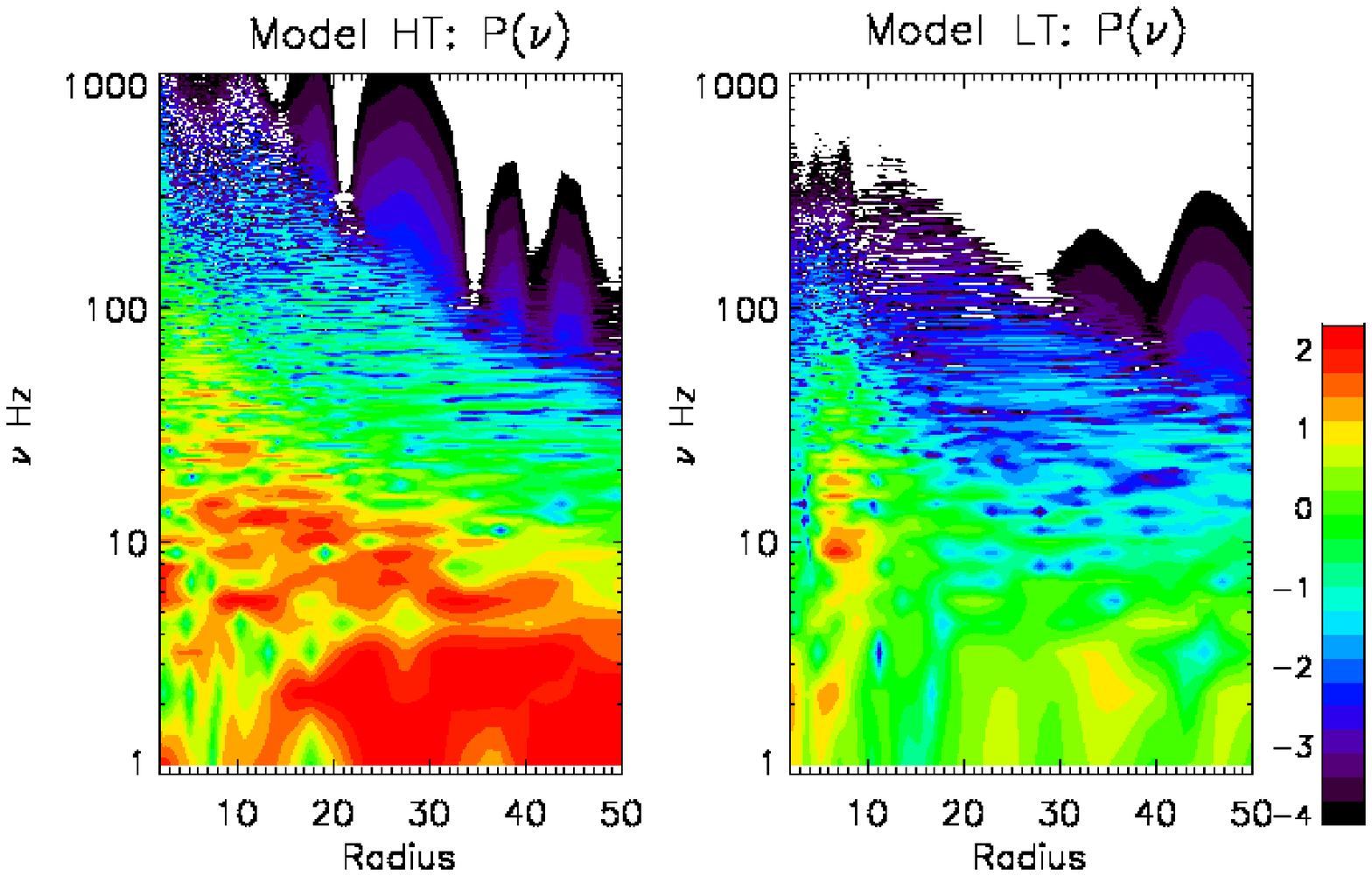}
 \end{center}
 \caption{Radial distribution of the PSD of the time 
variation of mass accretion rate 
for model HT measured in $23000 < t/t_0 < 32000$ (left)and 
for model LT measured in $55000 < t/t_0 <64000$ (right). 
 \label{fig:f16}}
\end{figure}

Figure \ref{fig:f16} shows the spatial distribution of the Fourier 
amplitude $\nu P_{\nu}$ of time variabilities in mass accretion rate 
for model HT in $23000 < t/t_0 < 32000$, and  
for model LT in $55000 < t/t_0 < 64000$.
In model HT, various peaks appear at various radius. 
On the other hand, in model LT,  
low frequency QPOs around $10$Hz appear in 
$5 <\varpi/r_{\rm s} < 10$, where the inner torus is formed. 

\begin{figure}
 \begin{center}
   \FigureFile(120mm,90mm){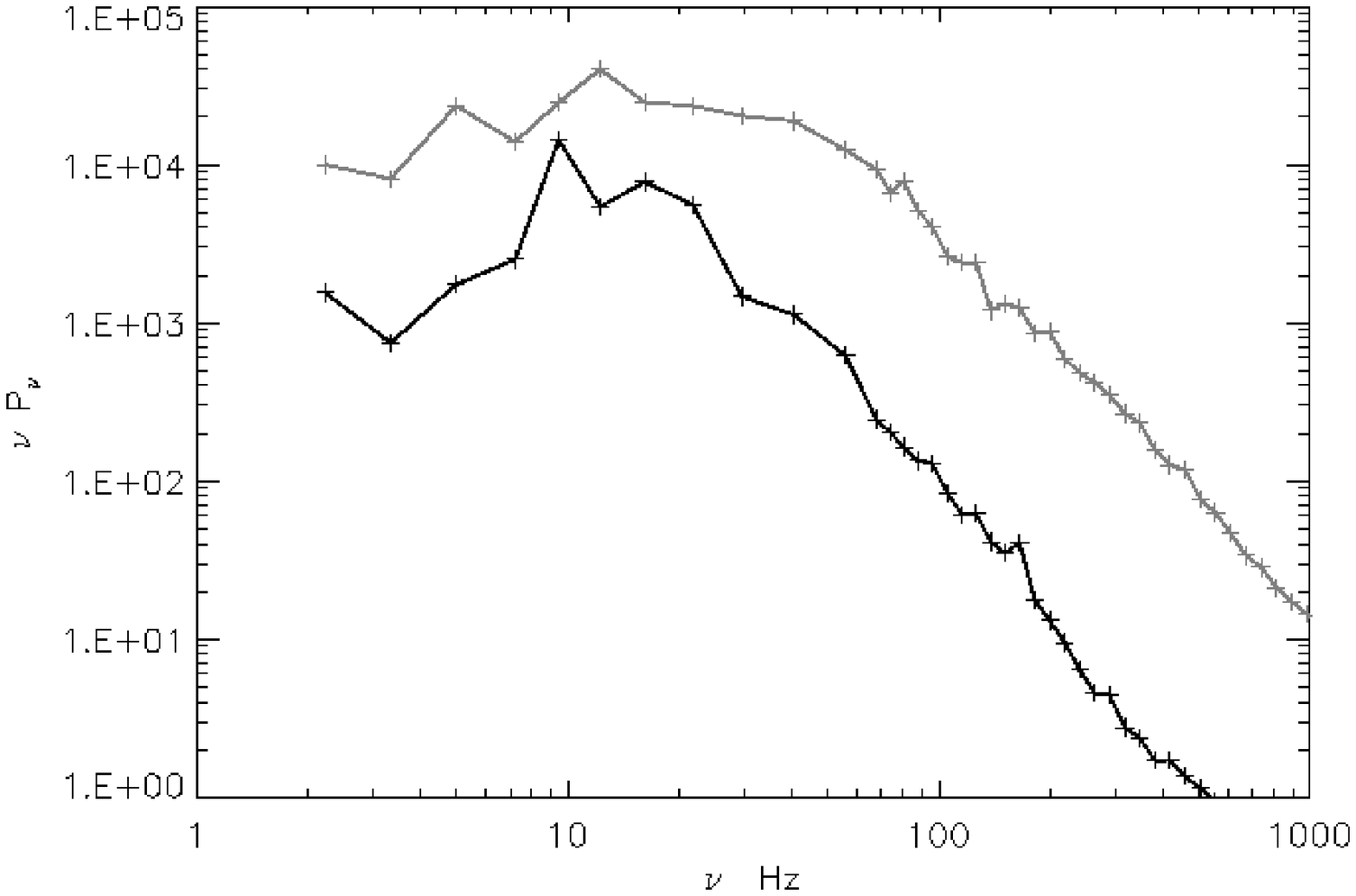}
  \end{center}
 \caption{ Power spectrum $\nu P_{\nu}$, where $P_{\nu}$ 
is the Fourier power of the time variation of mass accretion 
rate averaged in $2.5 < \varpi/r_{\rm s} < 29$ 
and $|z|/r_{\rm s} <1$ 
for model LT (black) and for model HT (gray). 
 \label{fig:f17}}
\end{figure}

Figure \ref{fig:f17} shows the 
Power Spectral Density (PSD) of the time variation of the mass accretion 
rate averaged in $3<\varpi/r_{\rm s}<8$ and $|z|/r_{\rm s}<1$. 
Black and gray curves show PSD ($\nu P_{\nu}$ where $P_{\nu}$ 
is the Fourier power) for model LT and model HT, respectively. 
We adapted the data in the time range 
$23000 < t/t_0 < 32000$ in model HT 
and $55000 < t/t_0 < 64000$ in model LT.

The PSD for model LT has a broad low-frequency peak 
around $10 {\rm Hz}$. 
This low-frequency peak corresponds to the oscillation 
involving amplification and release of magnetic energy in the inner torus. 
The PSD for model HT is flat in $\nu \leq 10{\rm Hz}$, 
which means that $P_{\nu} \propto \nu^{-1}$ 
and changes its slope around $\nu \sim 50 {\rm Hz}$. 
Since oscillations are excited at various radius in model HT, 
PSD shows flat spectrum. 
The PSD in model LT has a slope steeper than that in model HT 
in $30 {\rm Hz} < \nu <100 {\rm Hz}$.  

In model LT in Figure \ref{fig:f17}, 
a small peak appears at $\nu \sim 150 {\rm Hz}$. 
Based on the simulation using 32 azimuthal grid points, 
\citet{mat2007} pointed out that a high-frequency QPO 
appears when the disk shows low-frequency ($\sim 8 {\rm Hz}$) 
sawtooth-like oscillations of magnetic energy. 
Here we confirmed that low-frequency QPOs and high-frequency QPOs
coexist by a simulation including twice as many grid points 
in the azimuthal direction.

\section{Discussion} \label{discuss} 

In this paper, we studied the dependence of the structure and 
time variation of the black hole accretion flows on the 
gas temperature supplied from the outer region. 
When hot gas ($T \sim 10^{10} K$) is supplied, the angular 
momentum is transported efficiently all the way to the black hole. 
The average ratio of the Maxwell stress to gas pressure, 
$\alpha \equiv \langle B_{\varpi}B_{\varphi} /4 \pi \rangle /
\langle P \rangle \sim 0.05$, 
consistent with previous simulations by \citet{haw2001} 
and \citet{mac2003}. 
Such disks correspond to the radiatively inefficient, 
optically thin disk in the low/hard state of black hole candidates. 
Our simulation clearly showed the appearance of outflows with 
a maximum speed of $\sim 0.05c$ from such hot disks. 
The mass outflow rate $\dot{M}_{\rm out}$ is comparable to the 
mass accretion rate to the black hole.
%
The magnetic fields are turbulent inside the disk but 
they show large-scale coherent structures near the rotation axis. 
Plasma flows out in the interface between the funnel and 
the accretion disk.
The hot accretion flow shows time variations whose PSD 
is flat in $\nu \leq 10 {\rm Hz}$ when the mass of the 
central black hole is $M \sim 10 M_{\odot}$. 

When cool plasma is supplied, 
an inner torus is formed around $\varpi \sim 4-8 r_{\rm s}$. 
Such a torus is formed when the MRI-generated turbulent 
magnetic field is dissipated inside the disk. 
Since Maxwell stress decreases due to the decrease of magnetic 
energy, $\alpha$ decreases to $\alpha \leq 0.01$. 
Thus a nearly constant-angular-momentum torus is formed in the 
innermost region ($\varpi < 10 r_{\rm s}$) of the 
accretion flow. 
%
We found that the inner torus deforms itself into a crescent shape. 
Such deformation takes place due to the growth of the 
non-axisymmetric instability in geometrically thick tori 
\citep{pap1984}. 

Such a non-axisymmetric structure enhances the growth of MRI.
As magnetic energy increases, the angular momentum transport rate 
increases. 
Thus, the accretion rate increases. 
When the magnetic energy accumulated in the disk is released, 
the disk comes back to the weakly magnetized, 
axisymmetric torus. 
New cycle begins as magnetic energy is amplified by MRI. 
The period of the cycle is about $1000 t_0$ in model LT. 
It creates a low-frequency peak around $ 4-8 {\rm Hz}$ in 
PSD of the mass accretion rate. 
Low-frequency QPOs sometimes observed 
in low/hard state and hard intermediate state (HIMS) 
of black hole candidates can be 
reproduced by such magnetic cycles. 
We also showed that a high-frequency QPO is excited 
when low-frequency QPO appears. 

Let us discuss why the magnetic cycle is excited when low-temperature 
gas is supplied. 
When hot gas accretes, 
since the scale height of the gas is large, 
large eddies are formed in the magnetically turbulent disk and 
create a large scale coherent magnetic field 
which transports angular momentum efficiently. 
On the other hand, when low-temperature gas is supplied, 
since the scale height and eddy size become smaller,  
angular momentum is transported only in the local region. 
Thus the angular momentum is transported more efficiently in 
high temperature disks.

Low temperature plasmas can be supplied when cooling instability 
takes place in the outer disk. 
\citet{mac2006} carried out a global 3D MHD simulation of black hole 
accretion flows by including radiative cooling. 
They showed that when the density of the outer disk is sufficiently 
high, cooling instability takes place. 
The outer disk shrinks vertically, and 
forms a magnetically supported, cool, optically thin disk. 
When such cool plasma accretes, 
the magnetic cycle may be excited in the inner region.

The amplitude of the $m=1$ non-axisymmetric mode correlates 
positively with the Joule heating rate in model HT 
but anti-correlates in model LT. 
In model HT, since the angular momentum transport rate is 
large enough, mass smoothly accretes along the spiral channel, 
in which the magnetic fields are stretched and form a current sheet. 
Magnetic energy is released in such current sheets. 
Therefore, the amplitude of the $m=1$ mode 
correlates with the Joule heating rate. 
On the other hand, in model LT, non-axisymmetric $m=1$ 
pattern disappears when the magnetic energy is released.

\citet{hom2005a} pointed out that during the transition from the 
LHS to HSS, 
GX339$-$4 shows a sub-transition from HIMS dominated by power-law 
X-ray radiation to soft intermediated state (SIMS) 
dominated by radiation from an optically thick disk. 
They also showed that the low-frequency QPOs appear when the X-ray 
spectrum stays in HIMS and SIMS and sometimes in LHS. 
The high-frequency QPOs appear in the HIMS and SIMS. 
When the high-frequency QPO is observed, a low-frequency QPO is always 
observed in the X-ray spectrum. 
This tendency is consistent with our simulation results. 

A number of theoretical models have been proposed for 
high-frequency QPOs. 
\citet{abr2001} proposed  that high-frequency QPOs 
are formed by the resonance of radial and vertical oscillations.  
\citet{kats2001a} studied the excitation of high-frequency QPOs 
by resonance with the disk warp. 
In our simulation, a high-frequency peak appears around 
$\nu \sim 150 Hz$ when the low-frequency oscillation is prominent. 
In our simulation, however, 
the low-frequency oscillation was weakened  
due to the heating of the inner torus.
We expect that when the extra heating is extracted by 
radiative cooling, the magnetic cycle may continue  
and excite high-frequency QPOs. 
We would like to report the results of numerical simulations 
including radiative cooling in subsequent papers.

In this work, we treated the relativistic effect by using the 
pseudo-Newtonian potential. 
The accuracy of this approximation is worst 
near the black hole's horizon especially for spinning black holes. 
When we include the effects of black hole spin, 
the inner torus will be formed closer to the black hole. 
Thus, higher frequency QPOs will appear due to the oscillation of 
the inner torus. 
We should note that disk luminosity depends on the relativistic 
beaming and light bending. 
In subsequent papers, we would like to report PSDs 
obtained by relativistic ray-tracing of numerical results.

\bigskip

We are grateful to M.A. Abramowicz, S. Kato, W. Kul\'zniak, and 
M. Burusa for discussion. 
Numerical computations were carried out on VPP5000 at 
Center for Computational Astrophysics, CfCA of NAOJ (P.I. MM). 
This work is supported in part by Japan Society for the 
Promotion of Science (JSPS) Research Fellowships for 
Young Scientists (MM: 18-1907), and 
in part by Grants-in-Aid for Scientific Research of the Ministry of 
Education, Culture, Sports, Science, and Technology 
(RM: 17030003).


\end{document}